\documentclass[amsmath,preprint,tightenlines,aps,prd]{revtex4}

\def\be{\begin{equation}}
\def\ee{\end{equation}}
\def\bea{\begin{eqnarray}}
\def\eea{\end{eqnarray}}
\def\bml{\begin{subequations}}
\def\blea{\bml\begin{eqnarray}}
\def\elea{\end{eqnarray}\end{subequations}}

\begin{document}

\title{Can a circulating light beam produce a time machine?}
\author{Ken D. Olum}\email{kdo@cosmos.phy.tufts.edu}
\author{Allen Everett}\email{everett@cosmos.phy.tufts.edu}
\affiliation{Institute of
Cosmology, Department of Physics and Astronomy, Tufts University,
Medford, MA  02155}

\begin{abstract}
In a recent paper, Mallett found a solution of the Einstein equations
in which closed timelike curves (CTC's) are present in the empty space
outside an infinitely long cylinder of light moving in circular paths
around an axis.  Here we show that, for physically realistic energy
densities, the CTC's occur at distances from the axis greater than the
radius of the visible universe by an immense factor.  We then show
that Mallett's solution has a curvature singularity on the axis, even
in the case where the intensity of the light vanishes.  Thus it is not
the solution one would get by starting with Minkowski space and
establishing a cylinder of light.

\end{abstract}

\maketitle

Mallett, in a recent article \cite{Mallett}, wrote down a static,
cylindrically symmetric solution to the Einstein field equations of
General Relativity in the space surrounding an infinitely long axially
symmetric circulating cylinder of light with radiation energy density
$\epsilon$.  We take the $z$-axis to be the symmetry axis of the
cylinder. The solution is intended to approximate the solution in the
physically meaningful case of a light beam carried by a wave guide of
finite length spiraling around the $z$-axis.  Obviously one could at
best hope the approximation to be valid for $\rho < L$ where $L$ is
the length of the spiral along the $z$-axis and $\rho$ is the radial
distance from the axis of the cylinder, which we take to be equal to
the radius of the cylindrical region containing the spiral. The
solution obtained in \cite{Mallett} has the remarkable property that
in the region exterior to the cylinder a plane of constant $z$ contains
circular closed timelike curves (CTC's) enclosing the cylinder of all
radii $\rho >\rho_{\min } >\rho_0$, where $\rho_0$ is the radius of
the cylinder. The CTC's occur even if $\epsilon$ becomes arbitrarily
small, although in the limit $\epsilon \rightarrow 0$, $\rho_{\min }$,
the minimum radius of the CTC's, $\rightarrow \infty$. This result
seems unphysical, since it is difficult to believe that a
gravitational source of arbitrarily small mass could distort space so
drastically as to cause the formation of CTC's.

We first summarize the argument in \cite{Mallett}.  (Units 
are chosen such that $G =c = 1$, where $G$ is the gravitational constant.)
Mallett begins with an axially symmetric metric of the canonical form

\be\tag{I-5}
ds^2 = fdt^2 - 2wdtd\phi  - l d\phi ^2 - e^{\mu }(d\rho ^2 + dz^2)
\ee

[Throughout, equation number I-n refers to Eq.\ (n) in \cite{Mallett}.]  As a source in the
Einstein equation he inserts the energy momentum tensor of an axially
circulating light beam
\be\tag{I-2}\label{eqn:source}
T_{\mu \nu } = \epsilon \eta_{\mu }\eta_{\nu }
\ee
where $\epsilon$ is the radiation energy density,
\be
(x^0, x^1, x^2, x^3) = (t, \rho , z, \phi )\,,
\ee
\be\tag{I-3}
\eta_{\mu }\eta ^{\mu } = 0 
\ee
and
\be\tag{I-4}
\eta_{\mu } = (\eta_0, 0, 0, \eta_3) 
\ee

(Since $z$-independence is assumed, what one actually is talking about is an
infinitely long cylindrical surface carrying a circulating beam described by
its energy/unit length, $\mu$.)

Mallett then writes the Einstein equation corresponding to (I-2) and (I-3),
obtaining three different component equations (I-12)--(I-14). From these he
obtains the constraint equation
\be\tag{I-17}
\partial ^2\Delta /\partial \rho
^2 = 0
\ee
where
\be\tag{I-8}
\Delta ^2 = fl + w^2
\ee

Equation (I-17) has the solutions $\Delta =\rho$ and $\Delta
=$ constant.  Mallett chooses the solution $\Delta =\rho$ which
he says simplifies the field equations. He does not discuss this, but,
from Eq.\ (I-8) in the limit $w = 0$, this is consistent with the usual
Minkowski metric $f = 1$ and $l = \rho ^2$, with $\phi$ the usual
angular coordinate.  The other choice would lead to $f = l =$ constant and
\be\label{eqn:x3s}
x_3 = \rho \phi = s\,
\ee
a linear coordinate equal to the arc length $s$ along a circle
of radius $\rho$.

Next Mallett from (I-2), (I-3), and (I-4) obtains the result
\be\tag{I-19}
\eta ^0
= \xi \eta ^3
\ee
where
\be\tag{I-20}
\xi  = (w + \rho )/f
\ee

(The $\rho$ in (I-20) is actually $\Delta$ and would become a
constant if one took $\Delta=$  constant and $x^3 = s$.) He then
proceeds to solve the field equations using the \textit{ansatz} $\xi =
\eta ^0/\eta ^3 =$ constant. This would describe the usual
situation for an electromagnetic wave if $\Delta =$  constant, so that
$\eta ^3$ represented the density of the linear momentum $p^3$ in the
$x^3$-direction. However, if $\phi$ is an angular variable, the
choice of $\Delta = \rho$ means that, by analogy with Eq.\ 
(\ref{eqn:x3s}), $\eta^3$ is the density of $p^3$/$\rho$, and
since the energy and momentum densities in an electromagnetic wave are
equal, this would lead to $\xi = \eta ^0/\eta ^3 = \rho$
for values of $\rho$ for which the radiation density is
nonzero. Hence the combined \textit{ansatz} $\Delta =\rho$ and
$\xi =$  constant implies that the coordinates do not have their
conventional meaning.  Thus, e.\ g., the metric obtained in \cite{Mallett} in the
limit $\epsilon = w = 0$ has, from Eq.\ (I-20), the metric component $f =
g_{00}$ proportional to $\rho$ rather than the expected Minkowski
result $f = -1$.  One's first guess might be that the metric obtained in
\cite{Mallett} from the ansatz $\xi =$  constant differs from the metric of
conventional cylindrical coordinates only by a coordinate
transformation.  We will, however, see presently that the situation is
more complex than that.

Mallett proceeds by obtaining from the field equations in the region outside
the light beam, where the energy-momentum tensor $T_{\mu \nu }= 0$, the
differential equation
\be\tag{I-32}
\rho ^2\partial ^2w/\partial \rho ^2 - \rho \partial w/\partial \rho  + w = 0
\ee

This linear equation has the general solution $w = c_1\rho +c_2\rho \ln \rho$.  For the present case of interest this is
rewritten in \cite{Mallett} as 
\be\tag{I-33}
w = \lambda \rho  \ln(\rho /\alpha )
\ee
with $\lambda$ and $-\lambda \ln \alpha$ thus playing the role of the
two arbitrary constants in the solution of (I-32).  The parameter
$\lambda$ is an unspecified dimensionless constant proportional to
$\mu$, the radiation energy per unit length, and $\alpha$ is an
unspecified length.  The order of magnitude of both of these constants
can be obtained on dimensional grounds. The constant $\lambda$ must
depend on G as well as $\epsilon$, and hence we expect $\lambda$ to be
of the order of the dimensionless quantity $G\mu/c^4 (=\mu $ in our
units); i.\ e., $\lambda \approx$ the energy per unit length in the
natural units of Planck energy per Planck length, where $m_Pc^2=
c^2\sqrt{\hbar c/G}\approx$ 10$^{19}$GeV $\approx 10^9$J, and
$L_P= Gm_P/c^2\approx 10^{-35}$ m. As for $\alpha$, in the case of an infinite
cylinder, the only relevant geometrical quantity with the dimensions
of length is $\rho_0$, the radius of the light cylinder, so we expect
$\alpha \approx \rho_0$.

Taking $w$ to be given by (I-33) outside of the circulating light beam, i.\ e.,
for $\rho \geq \alpha$, Mallett obtains from the field equations together
with the constraint $\Delta ^2=\rho ^2$ 
\be\tag{I-36}
\bigskip l = \rho \alpha  [1 - \lambda \ln(\rho /\alpha )]
\ee
 
Eq.\ (I-36) implies that, for sufficiently large $\rho$, $l$ becomes
negative and hence $\phi$ becomes a timelike coordinate, so that
traversing a circle at sufficiently large $\rho$ allows one to
traverse a CTC and return to the same point in space at an earlier
time. Thus one has a time machine.  Note that the Cauchy
horizon enclosing the region containing CTC's extends to infinity and
is thus not compactly generated.  Thus the theorems of Tipler
\cite{tip76,tip77} and Hawking \cite{cpc} requiring violation of the
weak energy condition can be evaded, and we have CTC's without the
presence of regions of negative energy density.  (These theorems
would, however, rule out the creation of CTC's in any finite-size
approximation to this spacetime.)

However, even accepting the results of \cite{Mallett}, there is a
serious practical problem. CTC's occur for regions where $\ln
(\rho/\alpha) > 1/\lambda$, where $\alpha \approx \rho_0$, the radius
of the circulating light beam, and $\lambda \simeq \mu$. Suppose the
circulating light beam comes from the light from a laser of average
power P and beam radius r fed into a light pipe of the same diameter
wound in a tight spiral of radius $\rho_0$ around the $z$-axis. The
energy density in the laser beam is thus $\epsilon =P/(\pi r^2c)$, and
the energy per unit length is given by
\be\label{eqn:epsilon}
\mu =\frac{\pi P\rho_0}{cr}\,.
\ee

Let us take $P = 1$kW, $\rho_0 = 0.5$m and $r = 1$mm.  (The values of
$P$ and $\rho_0$ appear relatively realistic if perhaps a bit
optimistic.  But, as we will see, the specific numerical values are
essentially irrelevant, and could be changed by many orders of
magnitude without altering the conclusions.).  Putting $\lambda =
G\mu/c^4$, we obtain $\lambda =\pi GP\rho_0/(c^5r)$, which is of order
$10^{-46}$.  Thus, from (I-36), CTC's would only occur for $\ln
(\rho/\alpha) > 10^{46}$, or
\be
\rho > 10^{(10^{46})}\rho_0\,.
\ee
So, because of the logarithmic dependence in  (I-36), CTC's
are predicted to occur only outside a region whose radius is so
fantastically large that it cannot even be sensibly compared to the radius
of the visible universe

To be more realistic, one should remember that, in the physically
interesting case of a cylinder of circulating light of finite length $L$
the equations can at most be relevant only for $\rho < L$.  The
prediction of CTC's certainly cannot be regarded as reliable if they do not
occur in that range. For this to happen, the energy/unit length of the
light beam in natural units would be $1/N$, where $L/\rho_0 = N
>1 $, meaning the intensity of the light beam would have to be of the
order of that required to form a black hole unless the ratio of the
length to the width of the apparatus is huge. Note that the numerical
value of $\mu/c^2$,  from Eq.\ (\ref{eqn:epsilon}) and the numbers
given above, is about 10$^{-19}$ kg/m, emphasizing the extremely small
amount of mass in the rotating light cylinder in a realistic case, and
how unexpected it would be for such an object to give rise to any
noticeable distortion, much less the production of CTC's, in the
surrounding space.

From the above discussion there is no practical possibility of using an
apparatus with a circulating light beam to build a terrestrial time machine.
However, the discussion in the preceding paragraph might suggest that such an
apparatus could, in principle, lead to the production of CTC's. If true, this
would be of considerable significance, since the question as to whether CTC's
can ever be produced, even in principle, is of fundamental importance.
Therefore we should examine further what light, if any, is shed on this
question by \cite{Mallett}.

Unfortunately, it appears that the metric of \cite{Mallett} is not the metric
that one would get by starting from Minkowski space and establishing a
circulating cylinder of light.  It is true that it is almost
everywhere a solution to Einstein's equations with Eq.\
(\ref{eqn:source}) as a source, but at the origin $\rho = 0$ there is
a line singularity.  For example, the $trtr$ component of the Riemann
tensor is
\be
R_{rtr}^t = \frac{1}{8\rho^2}
\ee
and so diverges at the origin.  This is not a coordinate artifact, as
we can see by taking the scalar
\be
R^{\alpha\beta\gamma\delta} R_{\alpha\beta\gamma\delta}
=\frac{3}{4\alpha \rho^3}
\ee
which is also divergent.  Worse yet, this divergence has no dependence
on $\lambda$, and so persists if one takes the limit in which the
source intensity $\epsilon$ goes to 0.

Thus it appears that the metric of \cite{Mallett} describes a cylinder of
light circulating in a spacetime which is pathological even without
the light.  It is thus unlike the spacetime of van Stockum
\cite{vanStockum} studied by Tipler \cite{Tipler:ctc}, which does go
to Minkowski space as the source is removed.

Setting $\epsilon = 0$ and consequently $\lambda = 0$, we have the
metric
\be
ds^2 =(\rho/\alpha) dt^2-\rho\alpha
d\phi^2-\sqrt{\alpha/\rho}(d\rho^2+dz^2)
\ee
It is straightforward to compute the connection and curvature from this
metric and demonstrate the following properties.
\begin{enumerate}
\item Except for $\rho = 0$, it is a vacuum solution of Einstein's
equations.  However, it is not flat anywhere, and it has a curvature
singularity at $\rho = 0$.
\item The paths $\rho =$ constant, $z =$ constant, $d\phi/dt =
1/\alpha$ are null geodesics, so the light does not require any
external apparatus to keep it in circulation; the photonic crystals
discussed in \cite{Mallett} would not be necessary.  (In the van
Stockum/Tipler spacetime \cite{vanStockum,Tipler:ctc}, the dust is
kept in circulation by its gravitational attraction, but in the present case
the light is circulating on the geodesics of the background metric,
i.\ e., it is in orbit around the singularity.)
\item The length of the circle $z=$ constant, $\rho =$ constant is
$2\pi\sqrt{\rho\alpha}$, but a photon can traverse this loop in either
direction in time $2\pi\alpha$ independent of $\rho$.  Thus at large
distances the elapsed time is arbitrarily small as compared to the
distance traveled.  Therefore it is not surprising that a small
modification of this metric yields CTC's.
\end{enumerate}

Thus it appears that the closed timelike curves appearing in
\cite{Mallett} are the result of starting with a pathological
spacetime instead of Minkowski space.  There is no reason to believe
on the basis of \cite{Mallett} that CTC's could be produced in the
laboratory, even if we had sufficient technology to control a density
of electromagnetic radiation so large as to have measurable
gravitational effects.

K.D.O. was supported in part by the National Science Foundation.

\bibliographystyle{apsrev}
\bibliography{gr}

\end{document}